\documentclass[12pt,a4paper]{article}

\usepackage{graphics}
\usepackage{latexsym}
\usepackage{amsmath}
\usepackage{amssymb}
\usepackage{slashed}
\usepackage[usenames]{color}

\title{Why do galaxies have extended flat rotation curves?}
\author{Bruce Hoeneisen}
\date{\small{
Universidad San Francisco de Quito, Quito, Ecuador \\
Email: bhoeneisen@usfq.edu.ec \\
14 December 2024}
}

\begin{document}
\maketitle

\begin{abstract}
\noindent
Recent observations by Mistele \textit{et al.} show that the circular velocity curves of isolated
galaxies remain	flat out to the	largest	radii probed so	far, \textit{i.e.} $\approx 1$ Mpc.
The velocity decline beyond the expected virial radius is not observed. 
These results imply that the galaxy halo is in thermal equilibrium
even at large radii where particles did not have time to relax.
The galaxies must have already	formed in the isothermal state.
How is this possible?
In the present note we try to understand the formation of galaxies
with warm dark matter in the expanding universe.
\end{abstract}


\section{Introduction}

The present study is inspired by weak gravitational lensing
measurements that find galaxy circular velocities of test particles
to be approximately constant
out to the largest radii probed so far, \textit{i.e.} $\approx 1$ Mpc \cite{Lelli}.
The expected virial radius, beyond which the rotation velocity
should decline with the Kepler law, is not observed.
The flat rotation curves correspond to the ``isothermal sphere"
with density run $\rho(r) \propto r^{-2}$ with particles
with the Maxwell-Boltzmann distribution.
Consider a galaxy with a rotation velocity $V(r) = 200$ km/s.
The radius $r$ at which the rotation period equals the age
of the universe is $0.5$ Mpc. So the galaxy at large $r$ has
no time to relax to the isothermal equilibrium state:
the galaxy must have formed already in this isothermal state.
How is this possible?

\begin{figure}
\begin{center}
\scalebox{0.7}
{\includegraphics{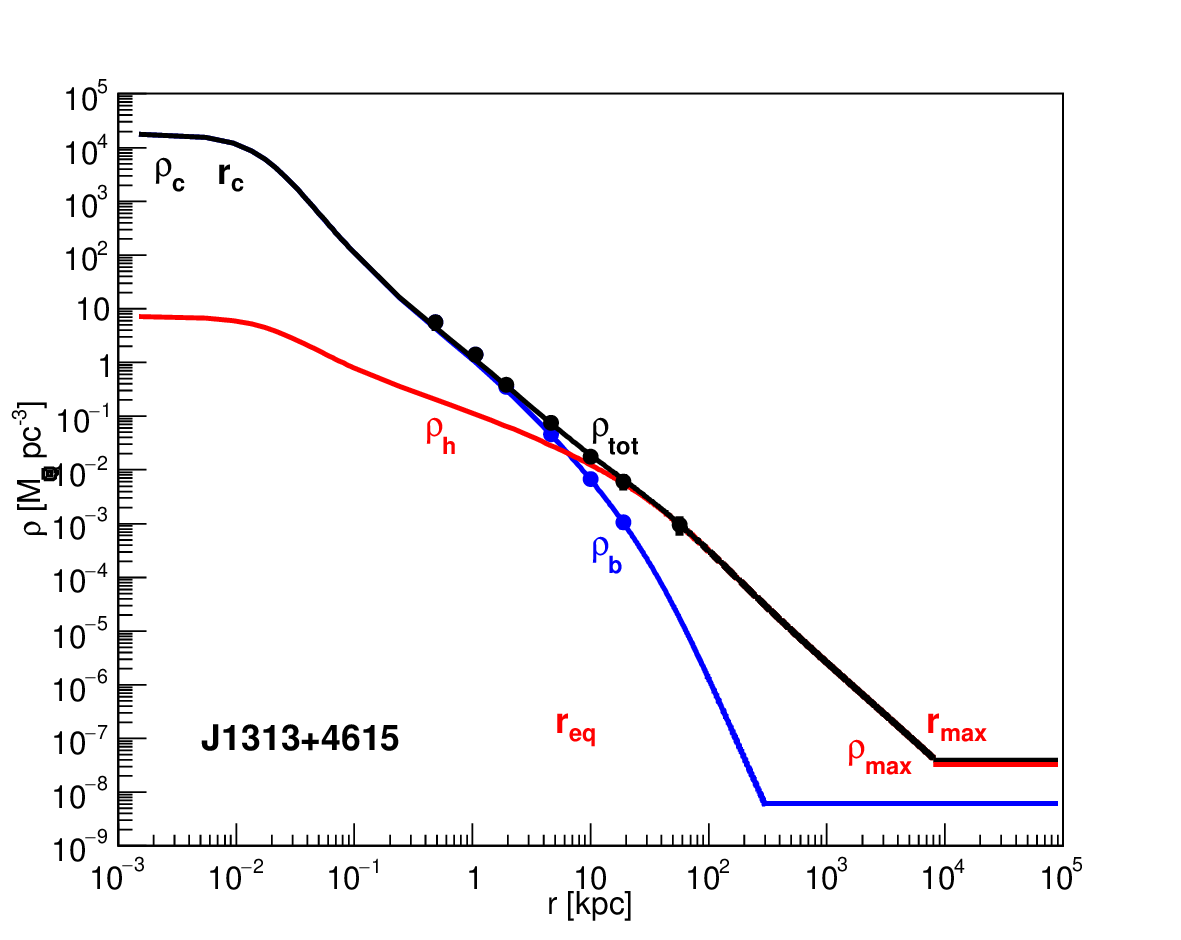}}
\caption{Observed \cite{Shajib} and calculated densities $\rho_\textrm{tot}(r)$, $\rho_b(r)$
and $\rho_h(r)$ of galaxy J1313+4615.
The fitted parameters are $\sqrt{\langle v_{rb}^2 \rangle}$,
$\sqrt{\langle v_{rh}^2 \rangle}$, $\rho_b(r_\textrm{min})$ and $\rho_h(r_\textrm{min})$
(or $v_{h\textrm{rms}}(1)$) \cite{understanding}.
Freeing a central black hole mass $M_{BH} = 0$ does not change the fit significantly.
The figure defines the core radius $r_c$ and density $\rho_c$, 
and the radius $r_\textrm{max}$ with density 
$\rho_\textrm{max} \equiv \rho_(r_\textrm{max}) = \Omega_m \rho_\textrm{crit}$
in this example (or a void, or the halo of a neighboring galaxy).
}
\label{J1313+4615_Fig_1}
\end{center}
\end{figure}

Figure 1 presents measured total densities $\rho_\textrm{tot}(r)$ and
baryon densities $\rho_b(r)$ of the large elliptical galaxy J1313$+$4615 \cite{Shajib}.
The dark matter density is $\rho_h(r) = \rho_\textrm{tot}(r) - \rho_b(r)$.
The curves are obtained by integrating numerically hydrostatic equations \cite{understanding}
that describe two self-gravitating classical non-relativistic gases,
``baryons" and ``warm dark matter", in mechanical and, separately, in thermal
equilibrium. 
To start the numerical integration it is necessary to provide four 
boundary conditions: the root-mean-square radial
thermal velocities $\sqrt{\left< v_{rb}^2 \right>}$ and
$\sqrt{\left< v_{rh}^2 \right>}$ (independent of $r$) and the core densities
$\rho_b(r_\textrm{min})$ and $\rho_h(r_\textrm{min})$, of baryons and of the dark matter halo,
respectively. These boundary conditions
are varied to minimize a $\chi^2$ between the numerical integration and the data.
``Baryons" are mostly neutral and ionized hydrogen and helium during the formation
of first generation galaxies, and mostly stars, dust and neutral and ionized gas 
in later galaxies.
The results of the fits are that the radial root-mean-square thermal velocities of dark matter particles
and baryons are similar, with 
\begin{equation}
\alpha \equiv \frac{\sqrt{\left< v_{rb}^2 \right>}}{\sqrt{\left< v_{rh}^2 \right>}}
\label{a}
\end{equation}
in the approximate range 0.5 to 0.7 in elliptical galaxies \cite{elliptical}.
(In spiral galaxies we find $\alpha$ in the approximate range 0.4 to 2.5 \cite{spiral}
as a result of galaxy rotation acquired, presumably, during galaxy collisions
and mergers. Rotating galaxies are beyond the scope of the present study.)
We note that baryons and warm dark matter have,
in general, different temperatures. Therefore non-gravitational 
dark matter-baryon interactions can be neglected on galactic scales.
Why is $\alpha$ of order 1?
At large radii $r > r_\textrm{eq}$ the dark matter
density dominates. At small $r < r_\textrm{eq}$ the baryon density dominates in 
large galaxies, while in dwarf galaxies dark matter may dominate even in the core. 
Excellent fits to the
observed density runs of baryons $\rho_b(r)$ and dark matter
$\rho_h(r)$ are obtained for dwarf \cite{dwarf}, spiral \cite{spiral} and
elliptical \cite{elliptical} galaxies with absolute luminosities
that span 4 orders of magnitude, and baryon core densities that span
6 orders of magnitude.
These excellent fits justify the hydrostatic equations 
with $\sqrt{\left< v_{rb}^2 \right>}$ and $\sqrt{\left< v_{rh}^2 \right>}$
independent of $r$.
If the galaxies have a third gas, \textit{e.g.} cold dark matter,
current observations can not distinguish it from the baryons
since we already obtain excellent fits to the data.
An important observation is that the warm dark matter core has an
adiabatic invariant $v_{h\textrm{rms}}(1)$ common to dwarf, spiral and elliptical galaxies, 
even though $\rho_h(r_\textrm{min})$ can be
orders of magnitude less than $\rho_b(r_\textrm{min})$, as in Figure \ref{J1313+4615_Fig_1}!
We interpret this adiabatic invariant to be of cosmological origin,
and identify $v_{h\textrm{rms}}(1)$ with the comoving 
root-mean-square thermal velocity of the non-relativistic warm dark matter particles
in the early universe (see section \ref{expanding_is} below).
There is no such observed adiabatic invariant for baryons, possibly       
because	baryons	have non-elastic collisions and radiate energy
(while the measured $v_{h\textrm{rms}}(1)$ for dark matter has a spread of a factor 3
between galaxies, the corresponding $v_{b\textrm{rms}}(1)$ for baryons
has a spread of $10^5$! \cite{elliptical} \cite{dwarf}).

The purpose of the present note is to try to understand Figure 1 and
the isothermal formation of galaxies. These studies are
a continuation of \cite{understanding} and \cite{elliptical}.

The basic building block of the galaxy is the isothermal sphere
that we briefly review in section \ref{cored_is}.
This isothermal sphere grows in thermal equilibrium due to the
expansion of the universe (section \ref{expanding_is}).
The dark matter core radius is determined by the dark matter ``warmness" 
$v_{h\textrm{rms}}(1)$ (section \ref{core}).
The preceeding results are valid even if the galaxy has a mix
of particles with different masses, so long as collisions
are elastic (section \ref{iso_vr}). The mix of warm dark matter
and baryons, and the effect of inelastic baryon collisions are
studied in section \ref{baryons}. Conclusions follow.

\section{The isothermal sphere}
\label{cored_is}

The flat rotation curves indicate that the galaxy approximates
an ``isothermal sphere" with particles that obey the
Maxwell-Boltzmann distribution \cite{understanding}.
For convenience we briefly review the isothermal sphere.
In the first approximation,
let us consider a galaxy as a self-gravitating
non-relativistic gas of warm dark matter particles of mass $m$.
These particles may be collisionless, or may collide elastically.
We are interested in spherically symmetric static solutions in
mechanical and thermal equilibrium.
The corresponding hydrostatic equations are Newton´s 
equation, and the equation of conservation of radial momentum:
\begin{equation}
\nabla \cdot \mathbf{g} = \frac{1}{r^2} \frac{d}{dr} (r^2 g_r) = -4 \pi G \rho, \qquad
\nabla P = \frac{dP}{dr} \mathbf{e}_r = \rho \mathbf{g}, \qquad
P \equiv \left< v^2_{r} \right> \rho,
\label{is_eq}
\end{equation}
where the mean-square dark matter particle radial velocity $\left< v^2_{r} \right>$
is independent of the radial coordinate $r$, \textit{i.e.} is isothermal.
The only solution with density run $\rho(r) \propto r^{-n}$ 
and $\left< v^2_{r} \right> \propto r^{-k}$ is
\begin{equation}
\rho(r) = \frac{\left< v^2_{r} \right>}{2 \pi G r^2}
\equiv \rho_c \left( \frac{r_c}{r} \right)^2
\label{is_sol}
\end{equation}
with $k = 0$.
Here we have defined $\rho_c r_c^2$.
The velocity of a test particle in a circular orbit is 
$V = \sqrt{2 \left< v^2_{r} \right>}$. Observed flat rotation curves at large $r$
indicate that $\left< v^2_{r} \right>$ is independent of $r$ and $\rho(r) \propto r^{-2}$,
\textit{i.e.} the galaxy halo is an isothermal sphere at large $r$.
The gravitational potential per unit mass with respect to a radial coordinate $r_c$ is
\begin{equation}
\Phi(r) \equiv -\int_{r_c}^r g_r dr = 2 \left< v^2_{r} \right> 
\ln{\left( \frac{r}{r_c} \right)}.
\end{equation}

The general solution of (\ref{is_eq}) depends on two boundary conditions,
\textit{i.e.} the core density $\rho_c$ at $r \rightarrow 0$, and $\sqrt{\left< v^2_{r} \right>}$
(and the mass of a central black hole that we will not consider here).
In other words, to initiate the numerical integration of (\ref{is_eq}) the
boundary conditions $\rho(r_\textrm{min})$ and $\sqrt{\left< v^2_{r} \right>}$ are required.
Here we will be interested in an approximate analytical solution
defined by two asymptotes: the density (\ref{is_sol}) for $r \gg r_c$, and
$\rho(r) = \rho_c$ for $r \ll r_c$. These two asymptotes meet at the 
core radius
\begin{equation}
r_c = \sqrt{ \frac{\left< v_r^2 \right>}{2 \pi G \rho_c}}.
\label{rc}
\end{equation}
Note that $\left< v_r^2 \right>$ is defined at large $r$.

The energy of one particle at $r$, with total momentum $p = m \sqrt{\beta v_r^2}$, is
\begin{equation}
E = \frac{1}{2} m \beta v_r^2 + 2 m \left< v_r^2 \right> \ln{\left( \frac{r}{r_c}\right)}.
\label{E}
\end{equation}
$\beta = 1$ if dark matter is collisionless and velocities are radial,
or $\beta = 3$ if dark matter particles have elastic collisions and the
velocities have become isotropic.

The mean number of particles ($< 1$) in a quantum state in the
non-degenerate gas is proportional to the Boltzmann factor
$\exp{\left( -E/kT \right)}$.
$k T$ is a constant (independent of $r$), with units Joule, called ``temperature".
The number of quantum states of a particle in the phase space volume $d^3\mathbf{r} d^3\mathbf{p}$
is proportional to this volume.
The number of particles per unit phase space volume is
\begin{equation}
\frac{dn}{d^3\mathbf{r} d^3\mathbf{p}} \propto \exp{\left( -\frac{E}{kT} \right)}.
\label{Liouville}
\end{equation}
The mean-square total velocity $\left< v^2 \right>$ 
obtained from (\ref{Liouville}) satisfies these equations:
\begin{equation}
\frac{1}{2} m \left< v^2 \right> = \frac{1}{2} m \beta \left< v_r^2 \right> = \frac{3}{2} k T, 
\label{kT}
\end{equation}
independently of $r$.
The energy of each particle in $dn$ is (\ref{E}). Since the exponential
in (\ref{Liouville}) separates into factors that depend either on $r$ or on $p$, the density
of the gas is $\propto r^{-2}$ if $k T = m \left< v_r^2 \right>$, so $\beta = 3$.
$\beta = 1$ is in disagreement with thermal equilibrium and the
observed flat rotation curves at large $r$.
We conclude that dark matter particles have elastic collisions, and velocities
become isotropic at least in the core.
In each volume element $d^3\mathbf{r}$, the particle velocities have the same Maxwell 
distribution with the same $k T$ and same $\left< v_r^2 \right>$ 
independent of $r$.

\section{The isothermal sphere in an expanding universe}
\label{expanding_is}

A homogeneous expanding universe has a matter density
\begin{equation}
\rho(a) = \frac{\Omega_m \rho_\textrm{crit}}{a^3},
\label{rho_a}
\end{equation}
where $a(t)$ is the expansion parameter (normalized to $a(t_0) = 1$
at the present time $t_0$).
We assume matter dominates so $a(t) \propto t^{2/3}$ $\propto H^{-2/3} \propto \rho^{-1/3}$.
The dark matter particle root-mean-square thermal velocity at expansion parameter $a$ is
\begin{equation}
v_{h\textrm{rms}}(a) = \frac{v_{h\textrm{rms}}(1)}{a}.
\label{vhrms1}
\end{equation}
$v_{h\textrm{rms}}(1)$ is the adiabatic invariant that defines 
how ``warm" the dark matter is.

Consider a positive density perturbation in a homogeneous 
expanding universe. An observer in this density peak ``sees"
dark matter expand adiabatically, reach maximum expansion, and then contract 
into the core of a galaxy.
By fitting galaxy rotation curves (or galaxy density runs)
it is possible to measure $\left< v^2_{r} \right> = V^2/2$
at large $r$, 
and $\rho_c = 3 (dV/dr)^2/(4 \pi G)$ at small $r$,
and obtain 
\begin{equation}
v'_{h\textrm{rms}}(1) = \sqrt{3 \left< v^2_r \right> }
\left( \frac{\Omega_c \rho_\textrm{crit}}{\rho_c} \right)^{1/3}.
\label{v'}
\end{equation}
If the expansion and contraction were free of relaxation and rotation,
$v'_{h\textrm{rms}}(1)$ would be equal to the adiabatic
invariant $v_{h\textrm{rms}}(1)$. However, due to relaxation and rotation,
in general $v'_{h\textrm{rms}}(1) \gtrsim v_{h\textrm{rms}}(1)$ \cite{measurements}.
The measured width of the distribution of $v'_{h\textrm{rms}}(1)$ determines
the contribution from relaxation and rotation (a factor $\gamma$ between 1 and $\approx 3$), 
and the lower bound of the measured distribution determines the adiabatic
invariant $v_{h\textrm{rms}}(1)$. Fits to dwarf galaxy rotation curves,
with a core density dominated by dark matter,
obtain $v_{h\textrm{rms}}(1) = 406 \pm 69$ m/s \cite{dwarf}.
A summary of \textit{measurements} that justify the interpretation that
$v_{h\textrm{rms}}(1)$ is of cosmological origin is presented in \cite{measurements}.

At $r > r_c$ the halo of the isolated galaxy approaches
the density run $\rho(r) \propto r^{-2}$ until it reaches, in our example, 
the mean density of the expanding universe (\ref{rho_a})
(or a void, or the halo of a neighboring galaxy). 
This behavior can
be seen by solving hydrodynamical equations \cite{understanding}.
The galaxy halo reaches $\rho(a_\textrm{max}) \equiv \rho_\textrm{max}$ at 
\begin{equation}
r_\textrm{max} = r_c \left( \frac{\rho_c}{\rho_\textrm{max}} \right)^{1/2}.
\label{r_max}
\end{equation}
At $a_\textrm{max}$, the Hubble expansion parameter is
\begin{equation}
H_\textrm{max} = H_0 \sqrt{\frac{\rho(a_\textrm{max})}{\rho_\textrm{crit}}}
= H_0 \sqrt{\Omega_m} a^{-3/2}_\textrm{max}.
\label{H_max}
\end{equation}
Note (from (\ref{rc}), (\ref{r_max}), (\ref{H_max}) and $\rho_\textrm{crit} = 3 H_0^2 /(8 \pi G)$) 
that the expansion velocity at $r_\textrm{max}$ is independent
of $a_\textrm{max}$:
\begin{equation}
H_\textrm{max} r_\textrm{max} \approx \sqrt{\frac{4}{3}} \sqrt{\left< v^2_r \right>}.
\label{H}
\end{equation}
The $\approx$ symbol is due to the inhomogeneity of the universe density
during galaxy formation. In (\ref{H}) we are neglecting the dark matter
thermal velocity 
$v_{h\textrm{rms}}(a_\textrm{max}) = v_{h\textrm{rms}}(1)/a_\textrm{max}$ 
at $a_\textrm{max}$.

The particles that are captured at $r_\textrm{max}$ by the growing galaxy halo 
form a galaxy in thermal equilibrium if
the expansion velocity $H_\textrm{max} r_\textrm{max} \approx \sqrt{3 \left< v^2_r \right>}$.
These particles populate the tail end of the Boltzmann distribution.
We note that $M(r) \propto r$, so the fraction of particles with energies
\begin{equation}
E = \frac{1}{2} m \cdot 3 \left< v^2_r \right> + 2 m \left< v^2_r \right> \ln{\left( \frac{r}{r_c} \right)}
\end{equation}
in the interval corresponding to $r$ and $r + dr$, is proportional to $dr$.

We also note that $r_\textrm{max}$ grows in proportion to $a_\textrm{max}^{3/2}$
while the separation between neighboring galaxies grows slower (in proportion
to $a_\textrm{max}$), so the universe becomes filled with galaxy halos 
leaving little intergalactic medium.

In conclusion, the halo formation is approximately 
isothermal without the need for relaxation:
the galaxy halo radius grows populating the tail of the Maxwell-Boltzmann 
distribution (\ref{Liouville}).

\section{The galaxy core}
\label{core}

Let us consider a dwarf galaxy with the core density
dominated by warm dark matter. We neglect dark matter particle
collisions during the first orbit.
A dark matter particle orbit has a distance of closest
approach to the galaxy center $r_\textrm{min}$ that is
obtained from $r_\textrm{max}$, the transverse thermal velocity 
$\approx \pm v_{h\textrm{rms}}(1)/a_\textrm{max}$ at $r_\textrm{max}$,
the velocity 
$\approx \sqrt{\left< v^2_r \right> \left( \beta + 4 \ln{(r_\textrm{max}/r_c)} \right)}$ 
in the core of a dark matter particle captured at $r_\textrm{max}$, 
and by conservation of angular momentum:
\begin{equation}
r_\textrm{min} = \pm r_c \frac{v_{h\textrm{rms}}(1)}{v'_{h\textrm{rms}}(1)} \cdot
f\left(\frac{\rho_c}{\rho_\textrm{max}} \right)^{1/3}.
\end{equation}
The function 
\begin{equation}
f\left(\frac{\rho_c}{\rho_\textrm{max}} \right) \equiv 
\left( \frac{\rho_c}{\rho_\textrm{max}} \right)^{1/6}
\frac{1}{\sqrt{\beta + 2 \ln(\rho_c / \rho_\textrm{max})}}
\end{equation}
lies in the range 0.5 to 1.4
for $\rho_\textrm{max} / \rho_c$ in the range $10$ to $10^5$,
and $\beta$ either 1 or 3.
So, the core radius $|r_\textrm{min}| \approx r_c$ implies that the
measured $v'_{h\textrm{rms}}(1)$ in the core of a galaxy
is approximately equal to adiabatic invariant $v_{h\textrm{rms}}(1)$
defined in (\ref{vhrms1}), and so
is indeed of cosmological origin (as argued in section \ref{expanding_is}
and in \cite{understanding},
and as confirmed by measurements summarized in \cite{measurements}).

\section{The iso-$\left< v_r^2 \right>$ sphere}
\label{iso_vr}

So far we have considered a gas of particles of mass $m$.
Let us now consider a gas with a mix of particles with different masses.
We still consider the case of particles that 
have elastic collisions. The results of sections \ref{cored_is} 
and \ref{expanding_is} remain 
valid, except that $E$ and $k T$ in (\ref{Liouville}) are proportional to the
particle masses, see (\ref{E}) and (\ref{kT}). 
The Maxwell-Boltzmann distribution of velocities remains unchanged
because $E/kT$ is independent of mass.
Note that, if particles are unable to exchange energy, particles
of different masses have different temperatures. However
in equilibrium
$\left< v_r^2 \right>$ remains the same for all particles,
independently of their mass, and
independent of $r$. In this case the ``isothermal sphere" should more
properly be called the ``iso-$\left< v_r^2 \right>$ sphere".
In the limit of baryons with elastic collisions, $\alpha = 1$ 
(with $\alpha$ defined in (\ref{a})).

\section{Adding baryons}
\label{baryons}

Let us consider the galaxy as a self-gravitating mix of two
gases: warm dark matter that has elastic collisions,
and baryons that have inelastic collisions. The hydrostatic equations
are two sets of equations like (\ref{is_eq}) separately 
for warm dark matter and baryons, with
$\mathbf{g} = \mathbf{g_h} + \mathbf{g_b}$ \cite{understanding}.
To start the numerical integration it is necessary to provide four
boundary conditions: $\sqrt{\left< v_{rb}^2 \right>}$,
$\sqrt{\left< v_{rh}^2 \right>}$,
$\rho_b(r_\textrm{min})$ and $\rho_h(r_\textrm{min})$.
These four parameters need to be taken from observations, predictions
or simulations.
Excellent fits to the data of dwarf, spiral and elliptical galaxies
justify taking $\sqrt{\left< v_{rb}^2 \right>}$ and
$\sqrt{\left< v_{rh}^2 \right>}$ independent of $r$.
The asymptotic solutions of the hydrostatic equations are:
\begin{eqnarray}
\rho_b(r): && \qquad \rho_b(r_\textrm{min}) \qquad 
\rightarrow \qquad \frac{\langle v_{rb}^2 \rangle}{2 \pi G r^2}
\qquad \rightarrow \qquad \propto \frac{1}{r^{2/\alpha^2}}, \label{rho_b} \\
\rho_h(r): && \qquad \rho_h(r_\textrm{min}) \qquad 
\rightarrow \qquad \propto \frac{1}{r^{2 \alpha^2}}
\qquad \rightarrow \qquad \frac{\langle v_{rh}^2 \rangle}{2 \pi G r^2}.
\label{rho_h}
\end{eqnarray}
$\alpha$ is defined in (\ref{a}).
These asymptotic solutions allow an understanding of Figure \ref{J1313+4615_Fig_1}. 
Note that in the limit $\alpha = 1$ we recover the ``iso-$\left< v_r^2 \right>$
sphere". Why is $\alpha < 1$ in elliptical galaxies \cite{elliptical}?
There are two reasons.
For first generation galaxies, the ``baryons", mostly hydrogen and helium,
become neutral and decouple from photons at redshift $z_\text{eq} =1090$, and so
for first generation galaxies
$v_{b\textrm{rms}}(1) \approx 8 \textrm{ m/s} \ll v_{h\textrm{rms}}(1) \approx 406 \textrm{ m/s}$.
Hydrodynamical equations show that the collapsing warmer
dark matter develops a core and forms later than the colder baryons \cite{understanding}.
The second reason for $\alpha < 1$ is that baryons have inelastic collisions
and gradually migrate towards the center of the galaxy halo.
In large galaxies, the core density is dominated by baryons.
The baryon core radius determines, and is equal to, the dark matter core radius.
The shrinking baryon core radius compresses the warm
dark matter in the core conserving the adiabatic invariant 
$v_{h\textrm{rms}}(1)$ (see Figure \ref{J1313+4615_Fig_1}).

Photometric and spectroscopic galaxy observations may obtain
the redshift $z$, the stellar mass $M_*$, and the baryon velocity dispersion
$\sqrt{\left< v_{rb}^2 \right>}$. If detailed density runs $\rho_b(r)$ are
observable, as in Figure \ref{J1313+4615_Fig_1}, 
then the break radius $r_\textrm{eq}$ is obtained, and a 
redundant measurement of $\sqrt{\left< v_{rb}^2 \right>}$ is possible:
\begin{equation}
\rho_\textrm{eq} \approx \frac{\left< v_{rb}^2 \right>}{2 \pi G r^2_\textrm{eq}}.
\label{rho_eq}
\end{equation}
Integrating the asymptotes (\ref{rho_b}) obtains
\begin{equation}
M_* = \frac{2 - 2 \alpha^2}{2 - 3 \alpha^2 } \cdot
\frac{2 \left< v_{rb}^2 \right>}{G} r_\textrm{eq}
\label{Mstar}
\end{equation}
(valid for $\alpha^2 < 2/3$). 
As a first approximation we may take $\alpha \approx 0.6$,
so (\ref{Mstar}) is another constraint between $r_\textrm{eq}$
and $\left< v_{rb}^2 \right>$.

The adiabatic invariant places another constraint between
the boundary conditions:
\begin{equation}
v'_{h\textrm{rms}}(1) = \sqrt{3 \left< v^2_{rh} \right> }
\left( \frac{\Omega_c \rho_\textrm{crit}}{\rho_h(r_\textrm{min})} \right)^{1/3}
\equiv \gamma v_{h\textrm{rms}}(1),
\label{v''}
\end{equation}
with $v_{h\textrm{rms}}(1) = 406 \pm 69$ m/s \cite{measurements},
and the relaxation factor $\gamma$ observed to be in the approximate range from 1 to 3.

Galaxy stellar masses $M_*$ may be related to primordial linear density
perturbations of total (dark matter plus baryon) mass $M$.
This mass $M$ is defined by the Press-Schechter
formalism with a gaussian window function 
and a power spectrum $P(k) \tau^2(k)$ with a cut-off factor $\tau^2(k)$ due
to the warm dark matter free-streaming \cite{PS} \cite{JWST} \cite{reionization}.
These Press-Schechter predictions, or their ellipsoidal collapse extensions 
pioneered by R.K. Sheth and G. Tormen \cite{Sheth_Tormen} \cite{Sheth_Mo_Tormen},
are in excellent agreement with galaxy stellar mass $M_*$ and ultra-violet
luminosity distributions in a wide range of redshifts \cite{JWST}.
Comparing these predictions with observations we obtain the following approximate 
relation \cite{JWST}:
\begin{equation}
\frac{M}{M_\odot} \approx 10^{1.5} \frac{M_*}{M_\odot}.       
\end{equation}

An empirical constraint between $M_*$ and $V$ is the baryonic 
Tully-Fisher relation for isolated galaxies \cite{Lelli}.
Similar relations are obtained from (\ref{rho_eq}) and (\ref{Mstar}):
\begin{equation}
M_* \propto V^2 r_\textrm{eq} \propto \frac{V^3}{\sqrt{\rho_\textrm{eq}}}.
\end{equation}

\section{Conclusions}

The observed extended flat rotation curves of galaxies \cite{Lelli} indicate
that galaxies are approximately isothermal spheres with
particles obeying the Maxwell-Boltzmann distribution \cite{understanding}.
The dark matter particles are collisional and these collisions are elastic.
The galaxies do not have time to relax to the isothermal equilibrium
state, so must have formed already in this state.
This isothermal formation is due to the growing galaxy halo
with density run $\rho(r) \propto r^{-2}$ at large $r$, 
and the expansion of the universe.
The particles falling into the growing galaxy halo potential well populate
the tail end of the Maxwell-Boltzmann distribution.
The halos grow until they meet voids or halos of neighboring galaxies.

The dark matter core radius is determined by the ``warmness" $v_{h\textrm{rms}}(1)$
of the dark matter. This adiabatic invariant is of cosmological
origin as shown by arguments in section \ref{expanding_is} and in \cite{understanding}, 
by measurements summarized in \cite{measurements},
and by the observed dwarf galaxy dark matter cores as shown in section \ref{core}. 
The measured warm dark matter
adiabatic invariant $v_{h\textrm{rms}}(1)$
happens to be in agreement with the ``no freeze-in and no freeze-out"
scenario of scalar dark matter coupled to the Higgs boson
\cite{measurements}.

``Baryons" have lower thermal velocities than dark matter
during the formation of first generation galaxies, and
have inelastic collisions, radiate
energy, and migrate
towards the bottom of the gravitational potential well, so 
$\alpha \equiv \sqrt{\left< v_{rb}^2 \right> / \left< v_{rh}^2 \right>}$ 
becomes less than 1.

In large galaxies, the core baryon density may dominate the
core warm dark matter density by several orders of magnitude,
as shown in Figure \ref{J1313+4615_Fig_1},
yet the measured adiabatic invariant in the
warm dark matter core remains invariant within uncertainties 
and relaxation corrections \cite{elliptical}.

Note that warm dark matter simulations should not neglect the
thermal velocity if the galaxy core is of interest.
If the intergalactic medium is of interest,
as in studies of the Lyman-$\alpha$ forest of quasar light,
it is necesary to cross-check that the simulations obtain
the observed extended galaxy halos with flat rotation curves \cite{Lelli},
since these halos leave little space to the ``intergalactic
medium".

\section*{Acknowledgements}

I thank Karsten M\"{u}ller for his early interest in this work and for
many useful discussions.


\begin{thebibliography}{7}

\bibitem{Lelli}
Mistele, T., McGaugh, S., Lelli, F., Schombert, J., Li, P. (2024)
Indefinitely Flat Circular Velocities and the Baryonic 
Tully-Fisher Relation from Weak Lensing.
arxiv:2406.09685

\bibitem{Shajib}
Shajib, A.J., Treu, T., Birrer, S., Sonnenfeld, A. (2021)
Dark Matter Halos of Massive Elliptical Galaxies at z = 0.2 are well
Described by the Navarro-Frenk-White Profile.
arxiv:2008.11724

\bibitem{understanding}
Hoeneisen, B. (2023) 
Understanding the Formation of Galaxies with Warm Dark Matter.
\textit{Journal of Modern Physics}, \textbf{14}, 1741-1754.

\bibitem{elliptical}
Hoeneisen, B. (2024)
Understanding Elliptical Galaxies with Warm Dark Matter,
\textit{Physics of the Dark Universe}, \textbf{46} (2024) 101643.

\bibitem{spiral}
Hoeneisen, B. (2019)
The Adiabatic Invariant of Dark Matter in Spiral Galaxies.
\textit{International Journal of Astronomy and Astrophysics}, \textbf{9}, 355-367.

\bibitem{dwarf}
Hoeneisen, B. (2022) Measurement of the Dark Matter
Velocity Dispersion with Dwarf Galaxy Rotation Curves.
\textit{International Journal of Astronomy and Astrophysics}, \textbf{12}, 363-381.

\bibitem{measurements}
Hoeneisen, B. (2024)
Measurements of the Dark Matter Mass, Temperature and Spin.
\textit{International Journal of Astronomy and Astrophysics}, \textbf{14}, 184-202.

\bibitem{PS}
Press, W.H. and Schechter, P. (1974) 
Formation of Galaxies and Clusters of Galaxies by Self-Similar Gravitational Condensation.
\textit{The Astrophysical Journal}, \textbf{187}, 425-438.

\bibitem{JWST}
Hoeneisen, B. (2024)
Are James Webb Space Telescope Observations Consistent with Warm Dark Matter?
\textit{International Journal of Astronomy and Astrophysics}, \textbf{14}, 45-60. \\

\bibitem{reionization}
Hoeneisen, B. (2022) 
Measurement of the Dark Matter Velocity Dispersion with Galaxy Stellar
Masses, UV Luminosities, and Reionization.
\textit{International Journal of Astronomy and Astrophysics}, \textbf{12}, 258-272.

\bibitem{Sheth_Tormen}
Sheth R.K., Tormen G., (1999)
Large-scale bias and the peak background split,
Mon. Not. R. Astron. Soc., \textbf{308}, 119-126.

\bibitem{Sheth_Mo_Tormen}
Sheth, R.K., Mo, H.J., Tormen, G. (2001)
Ellipsoidal collapse and an improved model for 
the number and spatial distribution of dark matter haloes,
Mon. Not. R. Astron. Soc. \textbf{323}, 1-12.

\end{thebibliography}
\end{document}